\documentclass[preprint,review,12pt]{elsarticle}

\usepackage{amssymb}
\usepackage{amsmath}

\usepackage[margin=2pt]{subcaption}
\usepackage{siunitx}
\usepackage{booktabs}
\usepackage[capitalize]{cleveref} %

\journal{Preprint}

\begin{document}

\begin{frontmatter}

\title{Mapping Social Media User Behaviors in Reciprocity Space} %

\author[ku]{Shiori Hironaka} %
\author[b]{Hayato Oshimo}
\author[c]{Mitsuo Yoshida}
\author[b]{Kyoji Umemura}

\affiliation[ku]{organization={Academic Center for Computing and Media Studies, Kyoto University},%
            addressline={hironaka@media.kyoto-u.ac.jp}, 
            state={Kyoto},
            country={Japan}}

\affiliation[b]{organization={Department of Computer Science and Engineering, Toyohashi University of Technology},
    state={Aichi},
    country={Japan}}
\affiliation[c]{organization={Institute of Business Sciences, University of Tsukuba},
    state={Tokyo},
    country={Japan}}

\begin{abstract}
Social media users exhibit diverse behavioral patterns as platforms function simultaneously as information and friendship networks. We introduce a reciprocity-based framework mapping users onto two-dimensional space defined by bidirectional connection ratios. Analyzing 48,830 Twitter users and 149 million connections, we demonstrate that fragmented user types from prior studies (influencers, lurkers, brokers, and follow-back accounts) emerge naturally as regions within continuous behavioral space rather than discrete categories. User properties vary smoothly across the reciprocity dimensions, revealing clear behavioral gradients. This framework provides the first unified model encompassing the full spectrum of social media behaviors and offers interpretable metrics for influence measurement and platform design.
\end{abstract}

\begin{keyword}
social media \sep user behavior \sep network analysis \sep reciprocity

\end{keyword}

\end{frontmatter}

\section{Introduction}

Social media platforms have fundamentally transformed human communication, creating complex ecosystems where users engage in diverse behavioral patterns~\citep{boyd2007,Jansen2009,Hanna2011,Chan-Olmsted2013}. These platforms uniquely function as both information networks and friendship networks, enabling users to simultaneously consume news, maintain social relationships, broadcast content, and seek information~\citep{Java2007,Krishnamurthy2008,Kwak2010,Myers2014,Romero2010}. This multifaceted nature produces remarkable behavioral diversity, yet understanding this diversity remains a central challenge in social media research.
A comprehensive framework that can systematically characterize the full spectrum of user behaviors across these interconnected functions would be highly beneficial, and researchers continue to pursue such unifying approaches.

Network structure provides an objective lens for understanding user behavior, complementing survey-~\citep{Brandtzaeg2009,Oh2015,Lin2011,Tominaga2018} and content-based~\citep{Kim2010} approaches.
Early network-based studies laid the foundation for structural approaches to user classification.
\citet{Java2007} analyzed Twitter's network topology and user activity patterns to identify distinct usage behaviors, revealing that users adapt platforms for fundamentally different purposes.
Building on this foundation, \citet{Krishnamurthy2008} pioneered systematic network-based categorization by examining follower versus following count relationships, identifying three broad user groups through in-degree versus out-degree analysis: broadcasters who maintain far more followers than people they follow (such as media outlets), acquaintances who maintain roughly reciprocal follower--following numbers (mirroring typical friendship networks), and evangelists or miscreants who follow many others but have relatively few followers (often including spammers or aggressive networkers).
This foundational classification revealed distinct behavioral clusters in the follower--followee space and established the principle that network structure can effectively capture different user intentions and behaviors.

As social media evolved, researchers discovered additional specialized types: \textit{influencers} broadcasting to large audiences~\citep{Cha2010,Bakshy2011,Huynh2019,Pagan2021,DeVerna2024,Harris2024}, \textit{lurkers} consuming content while rarely creating~\citep{Tagarelli2013,Edelmann2013,Tagarelli2014,Sun2014,Gong2015}, \textit{brokers} bridging distinct communities~\citep{Tsugawa2023}, and \textit{follow-back accounts} maintaining reciprocal connections~\citep{Hopcroft2011,Ghosh2012,Elmas2024}.
Each exhibits distinct network signatures, suggesting that network-based approaches have the potential to become universal measures of social media usage patterns.

Current network-based taxonomies face three fundamental limitations that impede a comprehensive understanding of user behavior.
First, they treat user types as discrete categories rather than continuous phenomena, forcing artificial boundaries on naturally fluid behaviors.
Second, existing metrics provide incomplete behavioral pictures: Followee-to-follower ratios are inherently one-dimensional, meaning two users with identical 1.0 ratios might have completely different connection patterns (one maintaining reciprocal friendships, another having separate follower and followee groups).
Third, discovered user types remain fragmented across studies without a unified theoretical framework, preventing systematic comparison and integration of findings.

These limitations stem from overlooking a critical network dimension~\citep{Almaatouq2016}---reciprocity, which represents the extent of bidirectional (i.e., mutual) connections between users~\citep{Garlaschelli2004,Squartini2013,Jiang2015,Ding2021,Gallo2025}.
While reciprocity is fundamental in network theory, existing classification schemes ignore how bidirectional connections fundamentally alter relationship meaning, distinguishing social bonds from information channels.
Although reciprocity is substantial on social platforms, Twitter exhibits relatively low reciprocity levels (ranging from 22\% to 58\% according to various studies~\citep{Java2007,Kwak2010,Rout2013,Myers2014}) compared to other online social networks~\citep{Kumar2006,Cha2009,Kumar2010}, indicating predominantly asymmetric relationships that existing one-dimensional metrics fail to capture adequately.
This oversight highlights the need to consider local reciprocity patterns at the individual user level, where the bidirectional connections vary significantly among users and reveal distinct behavioral patterns.

We propose a reciprocity-based framework to address these limitations. Our framework uses two complementary ratios: bidirectional edges to in-degree ($r_\mathrm{in}$) and bidirectional edges to out-degree ($r_\mathrm{out}$). This approach creates a continuous behavioral space where previously isolated user types emerge naturally as regions rather than discrete categories. Through analysis of 48,830 Twitter users and their complete ego networks comprising over 149 million connections, we conducted comprehensive behavioral characterization across multiple dimensions. We first identified four archetypal user categories at the extremes of the reciprocity space and analyzed their distinctive behavioral signatures through user property analysis, characteristic vocabulary extraction, and network formation patterns. We then examined the continuous nature of user behavior by mapping property variations across the entire reciprocity space, revealing smooth gradients rather than discrete boundaries between user types.

Our analysis demonstrates three key findings. First, fragmented user types from disparate studies unify systematically within our framework, with influencers, lurkers, brokers, and follow-back accounts occupying distinct reciprocity regions that correspond to the Feeding, Accumulating, Flowing, and Circulating archetypes, respectively. Second, user behavior varies smoothly across reciprocity space rather than exhibiting categorical boundaries, confirming that social media behaviors exist along continuous dimensions. Remarkably, we identify an intermediate \textit{high engagement zone} at moderate reciprocity values where content virality peaks, challenging assumptions that pure broadcasting or highly reciprocal relationships optimize engagement. Third, different user archetypes exhibit systematic connection formation patterns that reveal underlying network mechanisms; interaction preferences range from highly insular communities maintaining over 90\% within-group connections to information brokers bridging disparate network regions.
Our reciprocity-based framework is a promising approach that provides the first unified model encompassing the full spectrum of social media behaviors. It offers interpretable metrics for influence measurement, content recommendation, and platform design while illuminating how users navigate social media's multifaceted nature as both friendship and information networks.

\section{Defining the Reciprocity Space}

\begin{figure}[tp]
  \centering
  \includegraphics[width=.98\linewidth]{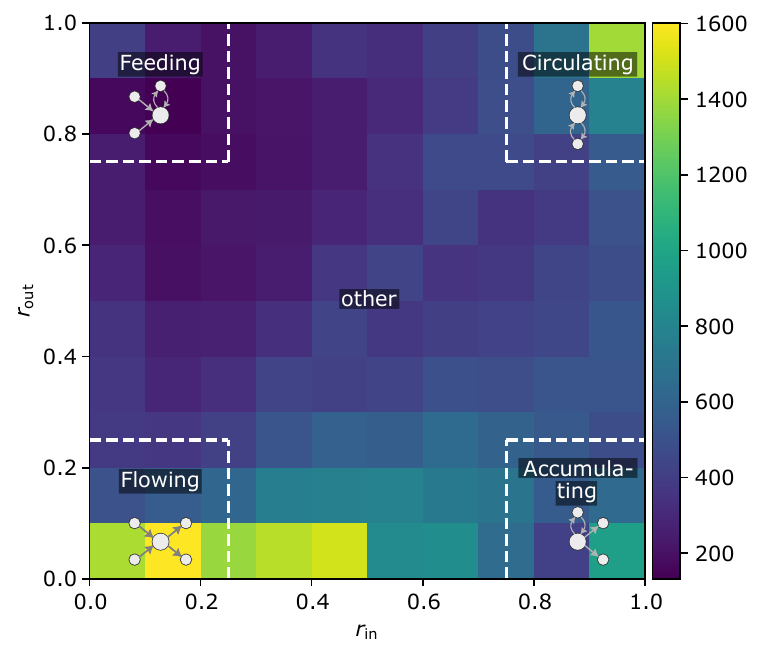}%
  \caption{User density distribution and archetype classification in reciprocity space. The two-dimensional reciprocity space shows the population distribution of 48,830 users, with $r_\mathrm{in}$ (bidirectional edges to in-degree) on the x-axis and $r_\mathrm{out}$ (bidirectional edges to out-degree) on the y-axis. Colors indicate user density, with four primary archetypes identified in the corner regions: Feeding (top left), Accumulating (bottom right), Flowing (bottom left), and Circulating (top right).}
  \label{fig:usercategory}
\end{figure}

We define a two-dimensional space based on the reciprocity of users' connections in directed social graphs.
For a given node, let $k_m$ denote the number of bidirectional connections (bidirectional edges), $k_i$ the in-degree (number of followers), and $k_o$ the out-degree (number of followees).
We define two reciprocity measures:
\begin{align}
  r_\mathrm{in} &= \frac{k_m + 1}{k_i + 1}, \\
  r_\mathrm{out} &= \frac{k_m + 1}{k_o + 1},
\end{align}
where the constant $+1$ avoids division by zero and can be interpreted as a self-edge. This ensures both ratios range from 0 to 1, with the minimum occurring when $k_m = 0$.
Here, $r_\mathrm{in}$ represents the proportion of reciprocal connections among incoming edges (followers), while $r_\mathrm{out}$ represents the proportion of reciprocal connections among outgoing edges (followees).
While these individual metrics have been used in previous studies~\citep{Gong2015,Takemura2015}, we are the first to combine them as coordinates in a two-dimensional behavioral space.

This two-dimensional ($r_\mathrm{in}$, $r_\mathrm{out}$) space provides greater expressiveness than the traditional followee-to-follower ratio $r_f$ alone.
The relationship between these metrics is as follows:
\begin{equation}
  r_f = \frac{k_o + 1}{k_i + 1} = \frac{r_\mathrm{in}}{r_\mathrm{out}}.
\end{equation}
Thus, users with similar followee-to-follower ratios can occupy different positions in our reciprocity space, capturing fundamentally different connection patterns.

\section{Data}

We collected tweets from Twitter's 1\% sample stream between July 11 and 13, 2021, extracting user IDs of those who posted at least one English tweet. After deduplication, we randomly selected 50,000 users. For each user, we retrieved up to 3,200 recent tweets and complete followee/follower information using Twitter's API between July 21 and 22, 2021. The final dataset included 48,830 users with complete data.

As a result, we collected 42,227,635 following relationships and 107,233,898 follower relationships. Merging these datasets yielded 149,368,679 unique following relationships, of which 21,482,872 were mutual (bidirectional) connections. The overall reciprocal relationship ratio among users connected by at least one edge in our constructed network was 16.8\%. This value is lower than those reported in other studies (21.1\% by \citet{Kwak2010} and 42\% by \citet{Myers2014}). 
This difference likely stems from our sampling methodology, as we sampled users appearing in Twitter's sample stream rather than employing snowball sampling or other network-based approaches.

\section{Behavioral Signatures based on User Archetypes}

\subsection{Archetype Classification and Analysis Methods}

To validate that the user reciprocity space captures meaningful behavioral differences, we identified users occupying the extreme corners of the reciprocity space, where the most pronounced behavioral patterns are expected to emerge.
We analyzed distinctive behavioral patterns for each user category based on user properties, characteristic vocabulary, and inter-category following relationships. User property analysis employed metrics across three dimensions: activity patterns, network structure, and content engagement.

Users were classified into four archetypes by dividing reciprocity into ranges using thresholds of 0.25 and 0.75 (\cref{fig:usercategory}).
These thresholds were chosen to isolate users with pronounced asymmetric ($\leq 0.25$) versus predominantly reciprocal ($\geq 0.75$) connection patterns while maintaining sufficient sample sizes for robust statistical analysis.
This choice is based on the principle that users at these extremes would exhibit the clearest manifestations of distinct behavioral archetypes, allowing us to characterize the fundamental patterns underlying the reciprocity space.
The resulting four archetypes represent the theoretical extremes of reciprocity behavior:
\begin{itemize}
\item Feeding class (low $r_\mathrm{in}$, high $r_\mathrm{out}$): These are users who reciprocate with most accounts they follow but attract many non-reciprocal followers, where $r_\mathrm{in} \leq 0.25$ and $r_\mathrm{out} \geq 0.75$, typical of content broadcasting patterns.
\item Accumulating class (high $r_\mathrm{in}$, low $r_\mathrm{out}$): These are users who maintain reciprocal relationships with most of their followers but follow many accounts without reciprocation, where $r_\mathrm{in} \geq 0.75$ and $r_\mathrm{out} \leq 0.25$, characteristics of information-seeking behavior.
\item Flowing class (low $r_\mathrm{in}$, low $r_\mathrm{out}$): These are users with predominantly unidirectional connections in both directions, where $r_\mathrm{in} \leq 0.25$ and $r_\mathrm{out} \leq 0.25$, representing minimal reciprocal engagement patterns.
\item Circulating class (high $r_\mathrm{in}$, high $r_\mathrm{out}$): These are users maintaining high reciprocity in both directions, where $r_\mathrm{in} \geq 0.75$ and $r_\mathrm{out} \geq 0.75$, reflecting strong bidirectional connection preferences.
\end{itemize}

From our sample of 48,830 users, we identified 5,640 Flowing users (11.6\%), 3,998 Accumulating users (8.2\%), 1,403 Feeding users (2.9\%), and 4,634 Circulating users (9.5\%).
The remaining 33,155 users (67.9\%) occupied intermediate positions in the reciprocity space, demonstrating the continuous nature of user behavior patterns that our framework captures.

For activity patterns, we analyzed tweet composition (original posts, retweets, replies, and quote tweets), posting frequency, total posts and likes, and account creation date. Tweet ratios were calculated from up to 3,200 collected tweets per user, with retweets, replies, and quote tweets treated as mutually exclusive categories, by removing retweets from other tweets. Daily tweet frequency was computed by dividing timeline tweet counts by the time span between the oldest and newest posts. We plotted the distribution of these metrics by using letter-value plot~\citep{Hofmann2017}.
Network structure was characterized using follower and followee counts extracted from user profile data.
Content engagement was measured through average retweeted and liked counts per tweet. To account for engagement decay, we calculated averages using only posts made before July 20, 2021 (24+ hours before data collection), applying original post timestamps for retweets. Metrics were computed both for all tweets and original content only.
To validate that these behavioral patterns represent distinct user populations, we performed Kruskal--Wallis tests~\citep{Kruskal1952} across all measured properties, followed by Conover's post-hoc test~\citep{conover1999practical} for pairwise comparisons between archetype pairs. To address the multiple testing problem, we applied Holm's method for $p$-value adjustment.

Characteristic vocabulary for each archetype was identified using chi-square statistics on English original tweets. Text preprocessing included removing mentions, URLs, special characters, emojis, and stopwords, followed by tokenization using NLTK's TweetTokenizer~\citep{Bird2009}. We calculated Chi-square values by treating each user's combined posts as a single document:
\begin{align}
E_{00} &= \frac{(N_{01} + N_{00})(N_{10} + N_{00})}{N_\mathrm{all}}, \\
E_{01} &= \frac{(N_{01} + N_{11})(N_{00} + N_{01})}{N_\mathrm{all}}, \\
E_{10} &= \frac{(N_{10} + N_{11})(N_{00} + N_{10})}{N_\mathrm{all}}, \\
E_{11} &= \frac{(N_{01} + N_{11})(N_{10} + N_{11})}{N_\mathrm{all}}, \\
\chi^2 &= \sum_{w \in \{0, 1\}} \sum_{u \in \{0, 1\}} \frac{(E_{wu} - N_{wu})^2}{E_{wu}},
\end{align}
where $N_{0*}$/$N_{1*}$ represent users not using/using a word, $N_{*0}$/$N_{*1}$ represent users outside/within the category, and $N_\mathrm{all}$ represents the total users.

Following relationships between archetypes were analyzed by counting cross-archetype connections and calculating row-normalized values (following tendency) and column-normalized values (follower tendency) to reveal interaction patterns between user categories.

\subsection{Behavioral Characteristics of Each Archetype}

\begin{figure*}[p]
    \centering
    \begin{subfigure}[t]{.24\linewidth}
        \centering%
        \includegraphics[width=.99\linewidth]{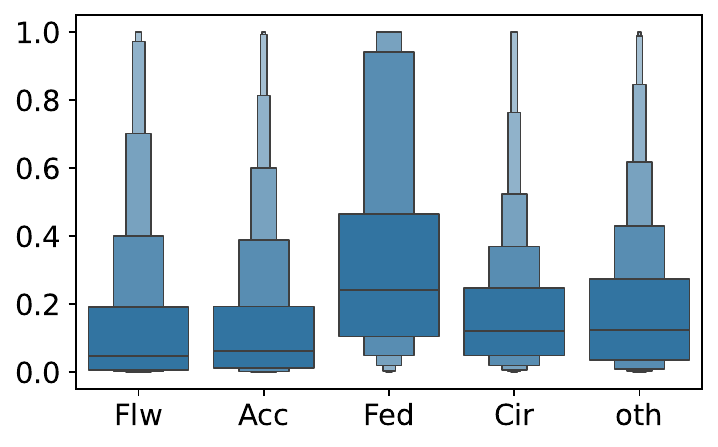}%
        \caption{Percentage of original posts} \label{fig:uproperty-p_original}%
    \end{subfigure}%
    \begin{subfigure}[t]{.24\linewidth}
        \centering%
        \includegraphics[width=.99\linewidth]{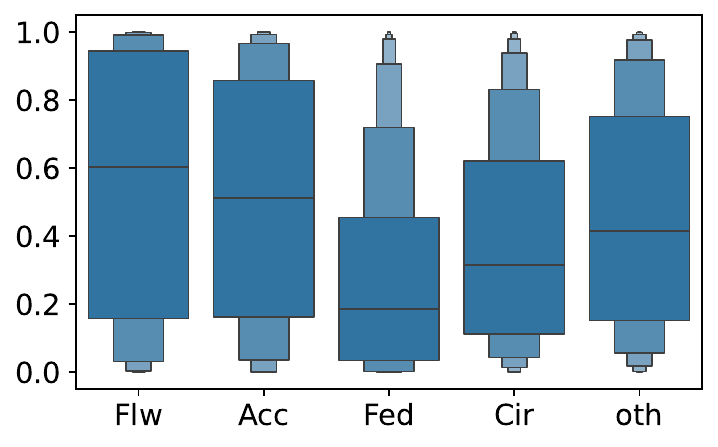}%
        \caption{Percentage of retweets} \label{fig:uproperty-p_retweets}%
    \end{subfigure}%
    \begin{subfigure}[t]{.24\linewidth}
        \centering%
        \includegraphics[width=.99\linewidth]{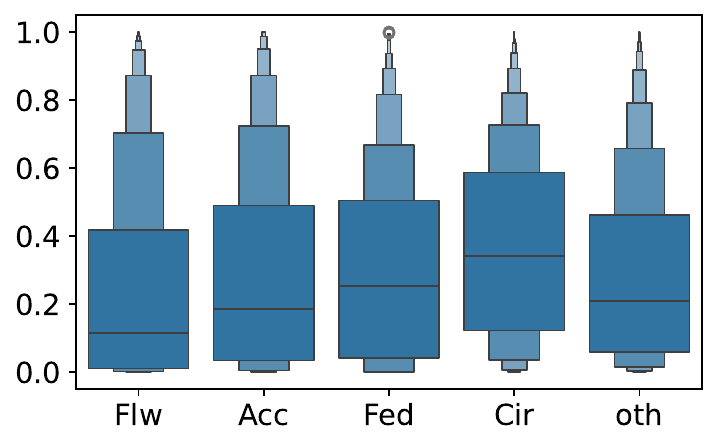}%
        \caption{Percentage of replies} \label{fig:uproperty-p_replies}%
    \end{subfigure}%
    \begin{subfigure}[t]{.24\linewidth}
        \centering%
        \includegraphics[width=.99\linewidth]{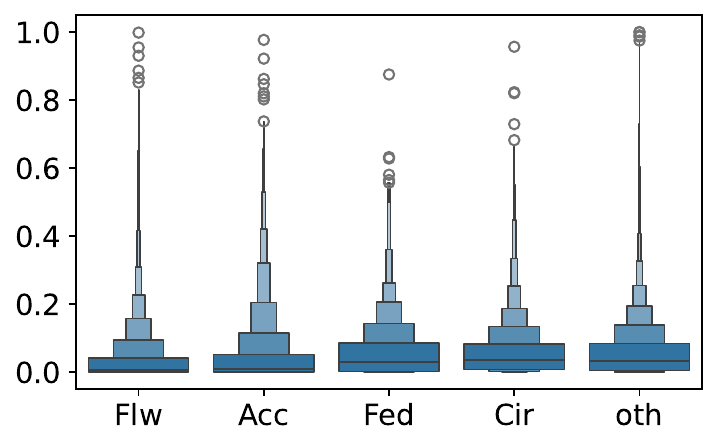}%
        \caption{Percentage of quote posts} \label{fig:uproperty-p_quotes}%
    \end{subfigure}\\
    \begin{subfigure}[t]{.24\linewidth}
        \centering%
        \includegraphics[width=.99\linewidth]{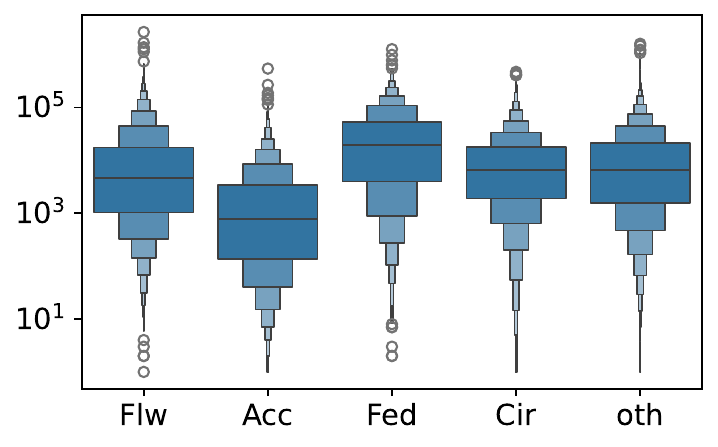}%
        \caption{Number of total posts} \label{fig:uproperty-statuses-count}%
    \end{subfigure}%
    \begin{subfigure}[t]{.24\linewidth}
        \centering%
        \includegraphics[width=.99\linewidth]{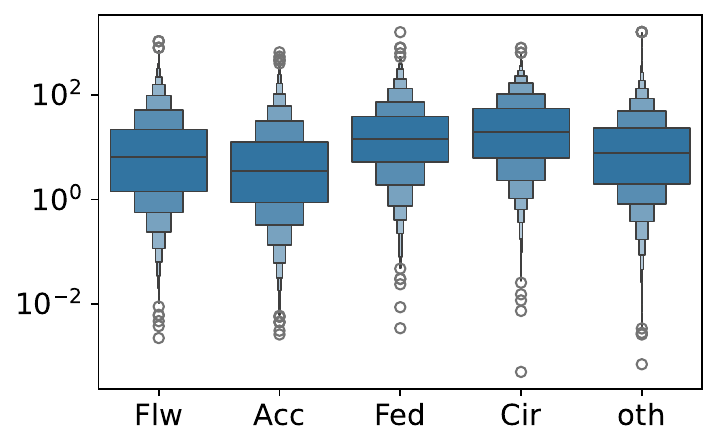}%
        \caption{Tweets per day} \label{fig:uproperty-tweets_per_day}%
    \end{subfigure}%
    \begin{subfigure}[t]{.24\linewidth}
        \centering%
        \includegraphics[width=.99\linewidth]{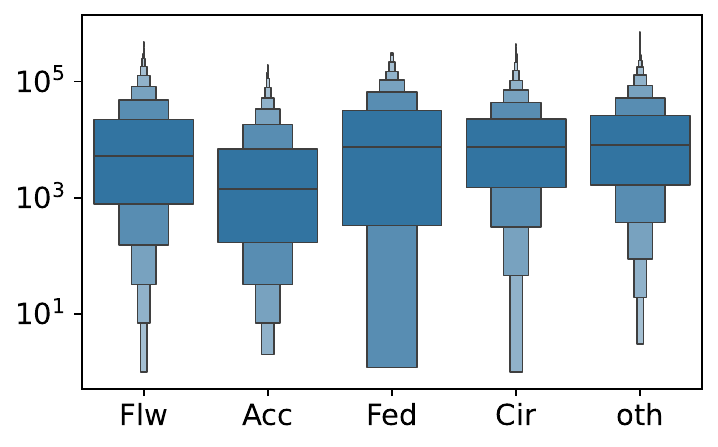}%
        \caption{Number of likes} \label{fig:uproperty-favourites_count}%
    \end{subfigure}%
    \begin{subfigure}[t]{.24\linewidth}
        \centering%
        \includegraphics[width=.99\linewidth]{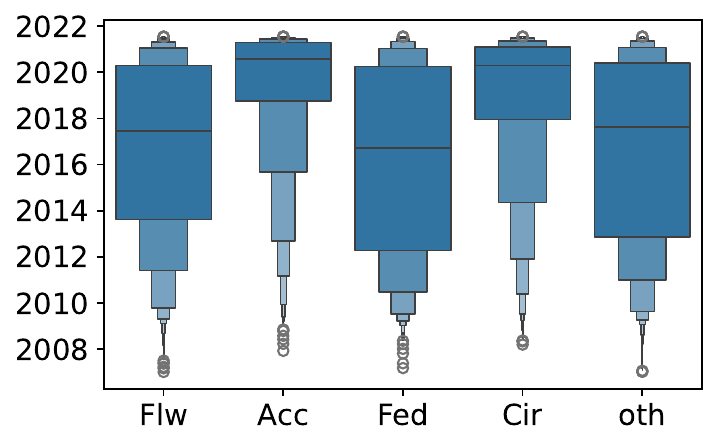}%
        \caption{Account creation date} \label{fig:uproperty-created_at}%
    \end{subfigure}\\
    \begin{subfigure}[t]{.24\linewidth}
        \centering%
        \includegraphics[width=.99\linewidth]{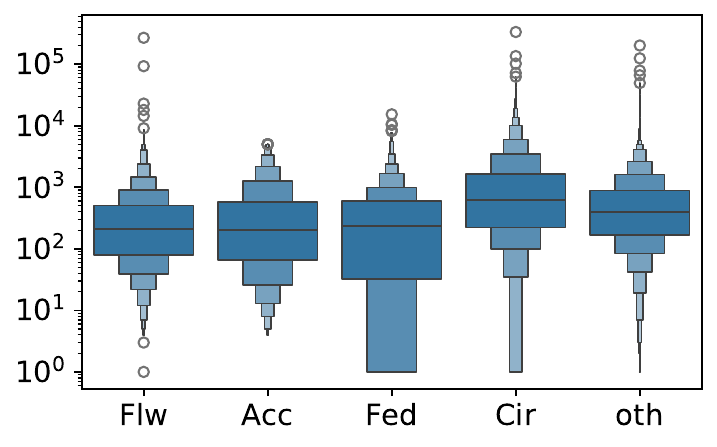}%
        \caption{Number of followees} \label{fig:uproperty-friends_count}%
    \end{subfigure}%
    \begin{subfigure}[t]{.24\linewidth}
        \centering%
        \includegraphics[width=.99\linewidth]{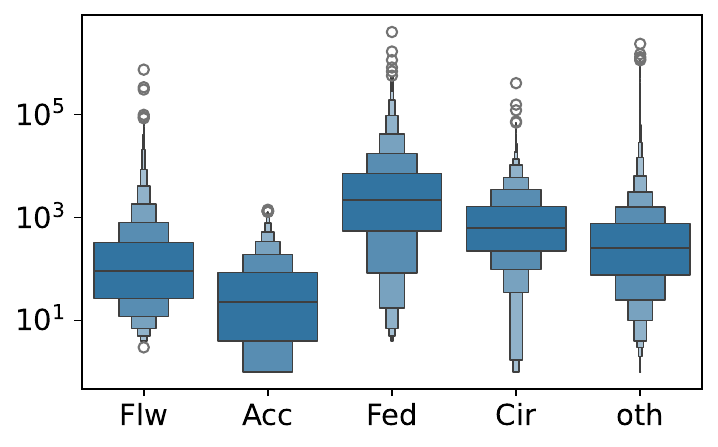}%
        \caption{Number of followers} \label{fig:uproperty-followers_count}%
    \end{subfigure}\\
    \begin{subfigure}[t]{.24\linewidth}
        \centering
        \includegraphics[width=.99\linewidth]{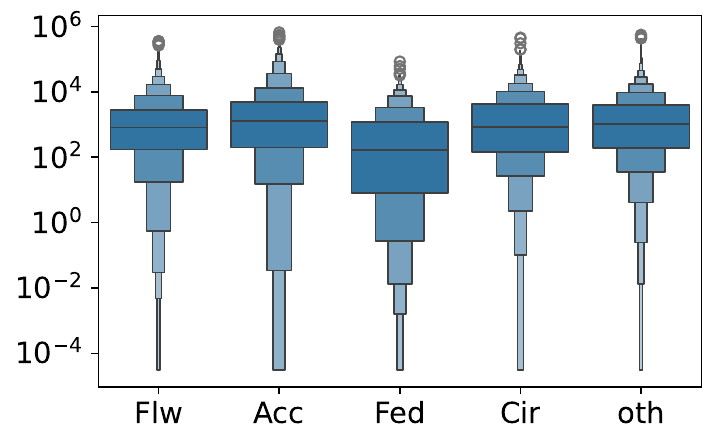}%
        \caption{Mean number of retweeted} \label{fig:uproperty-mean_retweeted}%
    \end{subfigure}%
    \begin{subfigure}[t]{.24\linewidth}
        \centering
        \includegraphics[width=.99\linewidth]{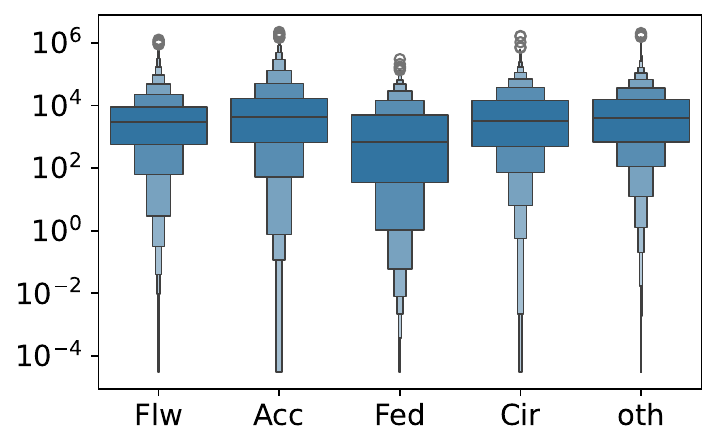}%
        \caption{Mean number of liked} \label{fig:uproperty-mean_favorited}%
    \end{subfigure}%
    \begin{subfigure}[t]{.24\linewidth}
        \centering
        \includegraphics[width=.99\linewidth]{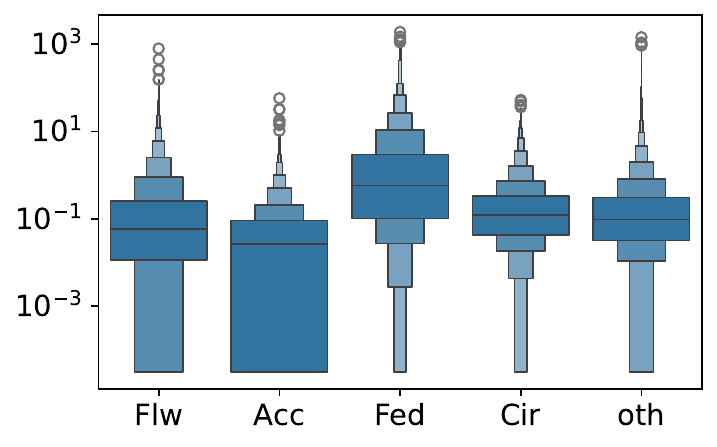}%
        \caption{Mean number of retweeted of original posts} \label{fig:uproperty-mean_original_retweeted}%
    \end{subfigure}%
    \begin{subfigure}[t]{.24\linewidth}
        \centering
        \includegraphics[width=.99\linewidth]{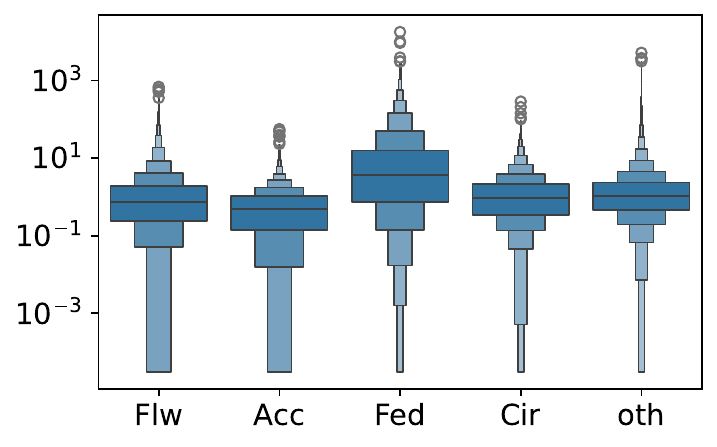}%
        \caption{Mean number of liked of original posts} \label{fig:uproperty-mean_original_favorited}%
    \end{subfigure}%
    \caption{\textbf{\scriptsize (For review: To be combined with \Cref{fig:behavioral-characteristics-b,fig:behavioral-characteristics-c} in final version.)} Behavioral signatures and network patterns of user archetypes. Letter-value plots show distributions of user properties across five categories: Flowing (Flw), Accumulating (Acc), Feeding (Fed), Circulating (Cir), and Other (Oth). Statistical significance across all behavioral metrics was confirmed through Kruskal--Wallis tests (all $p < 4 \times 10^{-124}$). (a--d) Tweet composition patterns showing percentage of original posts, retweets, replies, and quote tweets. (e--h) Activity and temporal metrics including total posts, daily posting frequency, total likes made, and account creation dates. (i--j) Network structure showing followee and follower counts. (k--n) Content engagement metrics measuring mean retweeted and liked for all posts and original posts specifically.}
    \label{fig:behavioral-characteristics-a}
\end{figure*}

\begin{figure}
    \centering
    \begin{subfigure}[t]{.49\linewidth}
        \centering%
        \includegraphics[width=.97\linewidth]{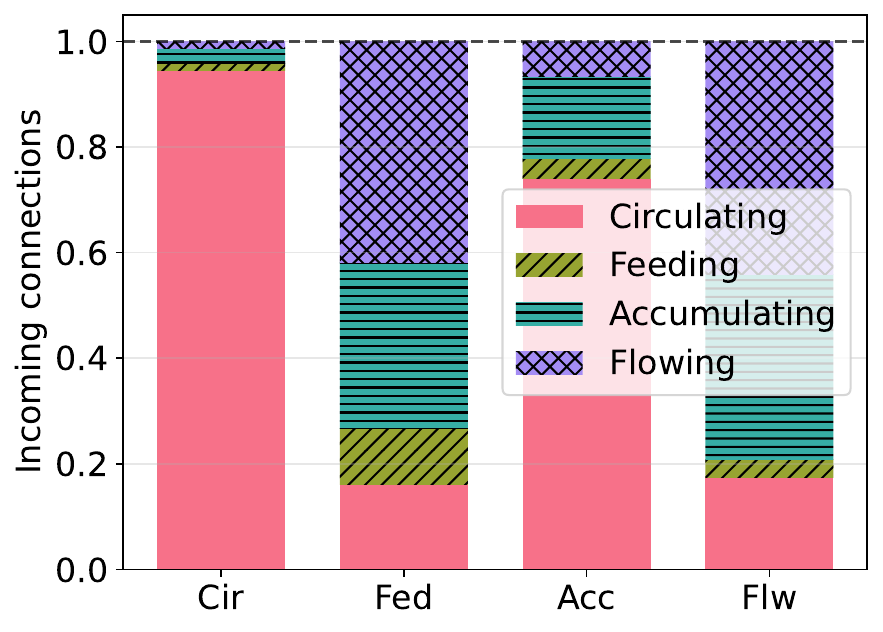}%
        \caption{Distribution of incoming follows by archetype} \label{fig:tendency-followers}%
    \end{subfigure}
    \begin{subfigure}[t]{.49\linewidth}
        \includegraphics[width=.97\linewidth]{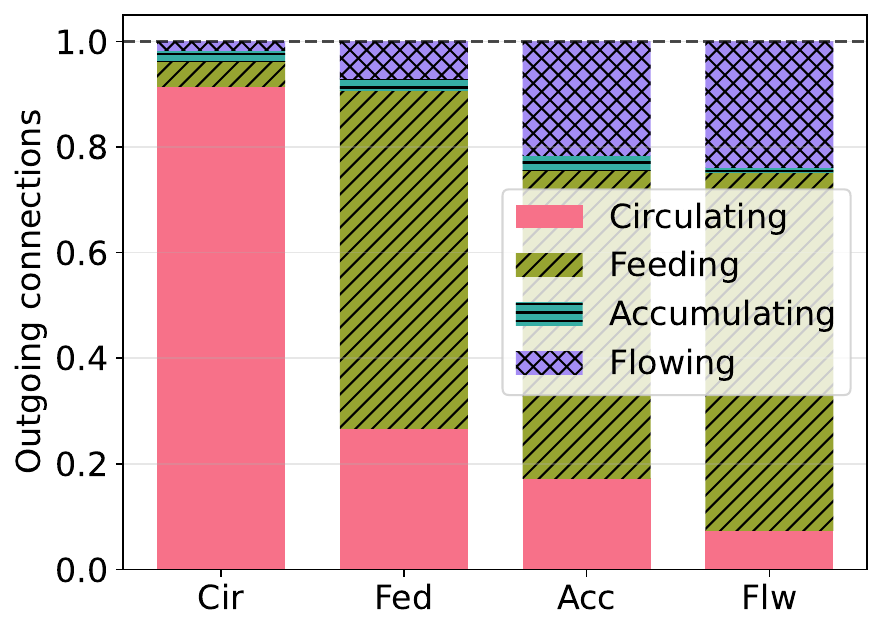}%
        \caption{Distribution of outgoing follows by archetype} \label{fig:tendency-following}%
    \end{subfigure}
    \caption{\textbf{\scriptsize (For review: To be combined with \Cref{fig:behavioral-characteristics-a,fig:behavioral-characteristics-b} in final version.)} Behavioral signatures and network patterns of user archetypes. (a--b) Following relationship patterns showing distribution of incoming and outgoing connections by archetype, revealing intergroup interaction preferences.}
    \label{fig:behavioral-characteristics-c}
\end{figure}

\Cref{fig:behavioral-characteristics-a,fig:behavioral-characteristics-b,fig:behavioral-characteristics-c} present comprehensive behavioral profiles for each user archetype across multiple analytical dimensions.
The letter-value plots (\Cref{fig:behavioral-characteristics-a}) reveal systematic differences in user properties across the four archetypes and intermediate users.
The characteristic vocabulary analysis (\Cref{fig:behavioral-characteristics-b}) identifies distinctive linguistic features for each archetype based on chi-square statistics, while the following relationship distributions (\Cref{fig:behavioral-characteristics-c}) reveal inter-archetype connection patterns.
These multifaceted results demonstrate that the reciprocity-based classification captures fundamentally different user behaviors that manifest consistently across diverse metrics.
We now examine each archetype's distinctive behavioral signature in detail.

Feeding users demonstrated information broadcasting behavior, characterized by significantly lower retweet rates (mean 29\%, $p < 4 \times 10^{-27}$) and higher original content production (mean 34\%, $p < 2 \times 10^{-60}$) compared to all other archetypes.
They maintained dramatically asymmetric networks, with significantly more followers ($\text{median}> \text{2,000}$) than followees (median 230) (\cref{fig:uproperty-friends_count,fig:uproperty-followers_count}), and achieved the highest engagement per original post (median 0.59 retweets, $p < 1 \times 10^{-81}$; 3.77 likes, $p < 1 \times 10^{-115}$) (\cref{fig:uproperty-mean_original_retweeted,fig:uproperty-mean_original_favorited}). Their vocabulary focused on scheduling and experiences (``tonight,'' ``date,'' ``weekend,'' and ``excited'') (\cref{tb:feature-words}). Despite showing moderate in-group preference (64.0\%), they attracted extensive following from other archetypes, particularly Flowing (67.8\%) and Accumulating (58.3\%) users (\cref{fig:tendency-following,fig:tendency-followers}). These patterns unify influencers and information broadcasters within our reciprocity framework. Unlike traditional approaches focusing solely on asymmetric audiences, our framework reveals that effective broadcasters strategically maintain high reciprocity in outgoing connections ($r_\mathrm{out} \geq 0.75$) while attracting predominantly non-reciprocal followers ($r_\mathrm{in} \leq 0.25$). This demonstrates strategic curation of followee networks alongside broadcasting to wider audiences, with evidence from \citet{Niitsuma2025} that content reposted by highly influential users reaches significantly broader audiences.

Accumulating users demonstrated clear information-seeking behavior with significantly higher retweet rates than Feeding and Circulating users (mean 51\%, $p < 8 \times 10^{-64}$ vs.\ Feeding and Circulating) (\cref{fig:uproperty-p_retweets}), lower activity levels than all other archetypes (median 769 total tweets, $p < 2 \times 10^{-307}$) (\cref{fig:uproperty-statuses-count}), and minimal follower counts (median 22, $p < 2 \times 10^{-230}$ vs.\ all other archetypes) (\cref{fig:uproperty-followers_count}). Their tweets received significantly less engagement than all other archetypes (mean retweets: $p < 4 \times 10^{-7}$; mean likes: $p < 2 \times 10^{-5}$) (\cref{fig:uproperty-mean_retweeted,fig:uproperty-mean_favorited}), and their vocabulary consisted of generic verbs and states (``stay,'' ``ready,'' ``drop,'' and ``rest'') (\cref{tb:feature-words}). They distributed follows across archetypes, particularly targeting Feeding users (58.3\%) (\cref{fig:tendency-following,fig:tendency-followers}).
These users unify previously scattered observations of lurkers and information seekers through their distinctive reciprocity pattern ($r_\mathrm{out} \leq 0.25$, $r_\mathrm{in} \geq 0.75$), which captures asymmetric consumption behavior: following many accounts while maintaining few reciprocal connections, indicating strategic information gathering rather than social relationship building.
While traditional lurker definitions focus on content consumption without creation, our reciprocity-based approach reveals that these users actively curate their information sources through strategic following patterns while maintaining minimal public engagement.
This aligns with characterizations of silent users~\citep{Gong2015} and demonstrates how reciprocity measures can quantitatively identify information consumption patterns.

Flowing users exhibited information brokering patterns with the highest retweet rates among all archetypes (mean 55\%, $p < 0.0008$) (\cref{fig:uproperty-p_retweets}). Their vocabulary mixed intimate terms with news-related content (``china,'' ``election,'' ``congress,'' and ``strategy'') (\cref{tb:feature-words}), and they demonstrated intermediate network formation patterns by extensively following Feeding users (67.8\%) while maintaining notable within-archetype connections (23.8\%) (\cref{fig:tendency-following,fig:tendency-followers}). These users unify network brokers within our reciprocity framework by maintaining minimal reciprocal connections in both directions ($r_\mathrm{in} \leq 0.25$, $r_\mathrm{out} \leq 0.25$). This captures the bridging behavior identified by \citet{Tsugawa2023} through more complex structural measures. Their mixed vocabulary confirms their role as information intermediaries spanning different domains, while low reciprocity patterns indicate that they connect disparate network regions without forming strong mutual bonds.

Circulating users exhibited the strongest interaction orientation, with significantly higher reply rates (median 34.0\%, $p < 1 \times 10^{-20}$) and quote tweet usage (mean 9.3\%, $p < 0.002$) than all other archetypes (\cref{fig:uproperty-p_replies,fig:uproperty-p_quotes}).
They maintained balanced followee and follower counts (medians 613 and 614), which differed significantly from other archetypes (number of followees: $p < 1 \times 10^{-86}$, number of followers: $p < 1 \times 10^{-39}$, vs.\ all other archetypes) (\cref{fig:uproperty-friends_count,fig:uproperty-followers_count}). These users formed highly insular communities, with 91.5\% of their follows and 94.5\% of their followers within the same archetype (\cref{fig:tendency-following,fig:tendency-followers}). Their characteristic vocabulary included intimate expressions (``moot,'' ``bestie,'' and ``ily'') and daily social interactions (``birthday,'' ``goodnight,'' and ``morning'') (\cref{tb:feature-words}).
These patterns align with the follow-back account phenomenon identified by \citet{Elmas2024}, although our framework reveals that this behavior spans both automated and human users, as evidenced by the combination of high reciprocity metrics and intimate vocabulary patterns.
\citet{Ghosh2012} demonstrated that such \textit{social capitalists} who follow back anyone who follows them are widespread on Twitter and often engage in link farming strategies to artificially inflate their follower counts.
Notably, this archetype presents an intriguing paradox: While exhibiting follow-back behaviors that theoretically promote openness, these users simultaneously form highly insular communities.
The extreme insularity combined with intimate vocabulary suggests that this archetype encompasses both genuine social communities seeking intimate communication through mutual following and coordinated accounts employing reciprocal strategies.
These findings that follow-back accounts maintain nearly 1:1 followee-to-follower ratios and tend to have a high number of followers align with our Circulating users' high reciprocity patterns.

While the four archetypes represent distinct behavioral patterns at the extremes of reciprocity space, a substantial portion of users (33,155 out of 48,830, representing 67.9\% of our sample) fell into the intermediate regions between these archetypes. Rather than exhibiting the pronounced characteristics that define the corner positions, these users display moderate reciprocity patterns that blend elements from multiple archetypes. This distribution demonstrates that the majority of social media users operate within a behavioral continuum rather than conforming to rigid categorical boundaries. The prevalence of intermediate users reinforces our framework's fundamental premise that user behavior exists along continuous dimensions. In this view, the four archetypes serve as reference points that anchor a broader spectrum of social media engagement patterns. These intermediate users represent the natural variation in how individuals balance information consumption, content creation, and relationship formation within the constraints and affordances of social media platforms.

\begin{table*}
        \caption{\textbf{\scriptsize (For review: To be combined with \Cref{fig:behavioral-characteristics-a,fig:behavioral-characteristics-c} in final version.)} Behavioral signatures and network patterns of user archetypes. Characteristic vocabulary table showing the top 20 words for each archetype with chi-square statistics indicating strength of association.}
        \centering
        \resizebox{\columnwidth}{!}{%
        \begin{tabular}{lSlSlSlS}
            \toprule
            Circulating & {$\chi^2$} & Feeding & {$\chi^2$} & Accumulating & {$\chi^2$} & Flowing & {$\chi^2$} \\ \midrule
            moot & 1299.8  & tonight & 427.1  & stay & 468.5  & bestie  & 649.2  \\
            bestie  & 1172.6  & dm  & 408.0  & morning & 433.0  & ily & 599.7  \\
            ily & 1039.3  & date & 342.8  & ready & 433.0 & youuu & 431.9  \\
            tl  & 690.9  & sent & 339.0  & dm  & 427.9  & moot & 423.1  \\
            youuu & 647.1  & weekend & 333.1  & drop & 423.1  & tl  & 400.4  \\
            okay & 622.1  & text & 333.1  & rest & 416.9  & rn  & 385.1  \\
            besties & 618.5  & saturday  & 328.2  & night & 414.1  & goodnight & 361.7  \\
            birthday  & 613.5  & room & 325.7  & close & 393.7  & besties & 326.1  \\
            goodnight & 604.4  & tomorrow  & 323.5  & view & 390.6  & okay & 309.7  \\
            aaa & 586.2  & send & 323.5  & break & 390.1  & china & 306.2  \\
            morning & 569.5  & excited & 316.9  & girl & 387.9  & twt & 293.4  \\
            twt & 539.1  & friday  & 315.4  & eye & 386.8  & profit  & 277.8  \\
            cute & 512.5  & later & 314.1  & full & 383.6  & election  & 271.5  \\
            rn  & 510.7  & gift & 312.8  & month & 382.8  & congress  & 262.3  \\
            dm  & 509.7  & taking  & 308.9  & coming  & 382.1  & im  & 259.1  \\
            sm  & 499.1  & inside  & 308.5  & waiting & 380.9  & idk & 251.2  \\
            rt  & 496.4  & putting & 306.9  & baby & 374.5  & fr  & 238.7  \\
            happy & 477.3  & set & 306.7  & mind & 371.5  & aaa & 237.7  \\
            love & 474.4  & promise & 305.7  & soon & 367.7  & launch  & 231.5  \\
            ofc & 468.1  & sunday  & 303.1  & cover & 367.2  & strategy  & 230.2  \\
            \bottomrule
        \end{tabular}%
        }
        \label{tb:feature-words}
    \label{fig:behavioral-characteristics-b}
\end{table*}

\section{Continuous Behavioral Gradients Across Reciprocity Space}

Having established four archetypal behaviors at the extremes of the reciprocity space, we now examine the continuous nature of the behavioral space where 67.9\% of users reside.
To visualize behavioral gradients, we divided the reciprocity space into a $10 \times 10$ grid. For each cell, we calculated median property values across all users within that reciprocity range and visualized results as heatmaps with contour lines indicating smooth transitions.
This analysis directly addresses the fundamental limitations of existing classification approaches. Unlike HITS scores~\citep{Kleinberg1999} that lack intuitive interpretation or simple followee-to-follower ratios that obscure behavioral distinctions, our reciprocity-based framework reveals the true continuous nature of user behavior that previous taxonomies failed to capture. The analysis included user metrics across activity patterns (tweet composition, posting frequency, and engagement), network structure (follower/followee counts), and content engagement for both all posts and original posts.

The results are shown in \cref{fig:user-property-continuous}.
Most properties showed smooth gradients across the space, confirming that the four archetypes represent extremes of continuous distributions rather than discrete categories.
This finding fundamentally challenges the rigid categorization assumptions underlying previous studies that treated user types as mutually exclusive groups.
Network structural properties displayed expected patterns: Followee count increased toward high values of both reciprocity measures, while follower count peaked at low $r_\mathrm{in}$ and high $r_\mathrm{out}$ (the Feeding region). Account age showed interesting variation, with newer accounts more prevalent in high-reciprocity regions, suggesting evolving platform usage patterns over time.

Remarkably, content engagement metrics (retweet and like counts) peaked not at the archetype extremes but in an intermediate high engagement zone (approximately $r_\mathrm{in}$ 0.5--0.7, $r_\mathrm{out}$ 0.6--0.9). This region, characterized by moderate reciprocity in both dimensions, appears optimal for content virality. Users in this zone balance sufficient reciprocal connections for initial content endorsement with enough asymmetric followers for broader reach.
This finding directly contradicts the limitations of one-dimensional approaches that would classify such users identically despite their strategic positioning for optimal engagement. 
The discovery of this high engagement zone exemplifies how our framework resolves the interpretational ambiguities that plague existing metrics. Two users with identical followee-to-follower ratios near 1.0, which traditional approaches would classify similarly, occupy distinct positions in our reciprocity space with markedly different engagement outcomes. This demonstrates that strategic positioning in intermediate reciprocity regions, rather than pure broadcasting or highly reciprocal relationships, optimizes content virality.

Our reciprocity-based approach reveals that specialized behaviors from prior studies (influencers, lurkers, brokers, and follow-back accounts) represent positions within a continuous behavioral space rather than discrete phenomena. This two-dimensional reciprocity space provides a generalizable framework applicable across social media platforms, unlike platform-specific methods or complex structural measures used previously. User types exist on a spectrum with smooth gradients in properties across reciprocity space, allowing mixed characteristics and transitions between types over time. The high engagement zone in intermediate reciprocity regions provides novel insights that challenge assumptions about rigid user categorization while offering practical guidance for content creators seeking optimal virality.

\begin{figure*}[tp]
    \centering
    \begin{subfigure}[t]{.24\linewidth}
        \centering%
        \includegraphics[width=.98\linewidth]{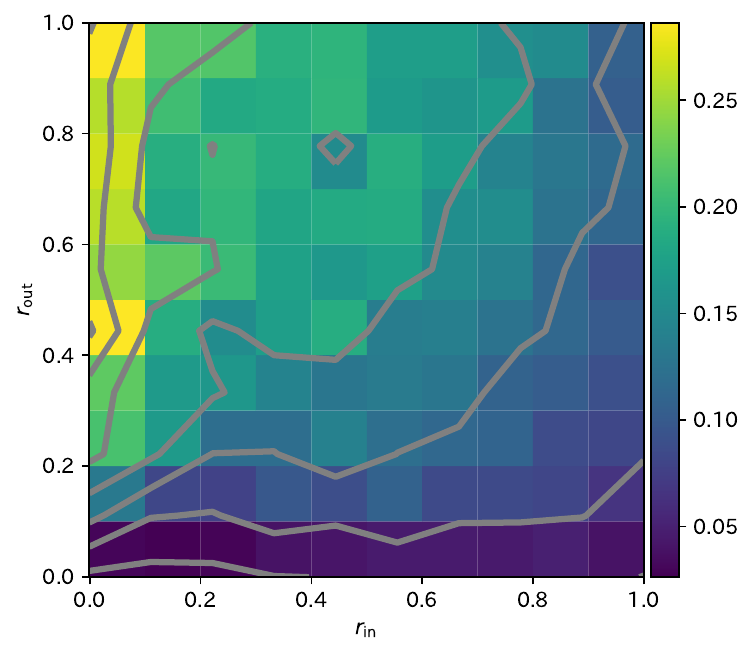}%
        \caption{Percentage of original posts} \label{fig:mutualratio-contour_p_original}%
    \end{subfigure}%
    \begin{subfigure}[t]{.24\linewidth}
        \centering%
        \includegraphics[width=.98\linewidth]{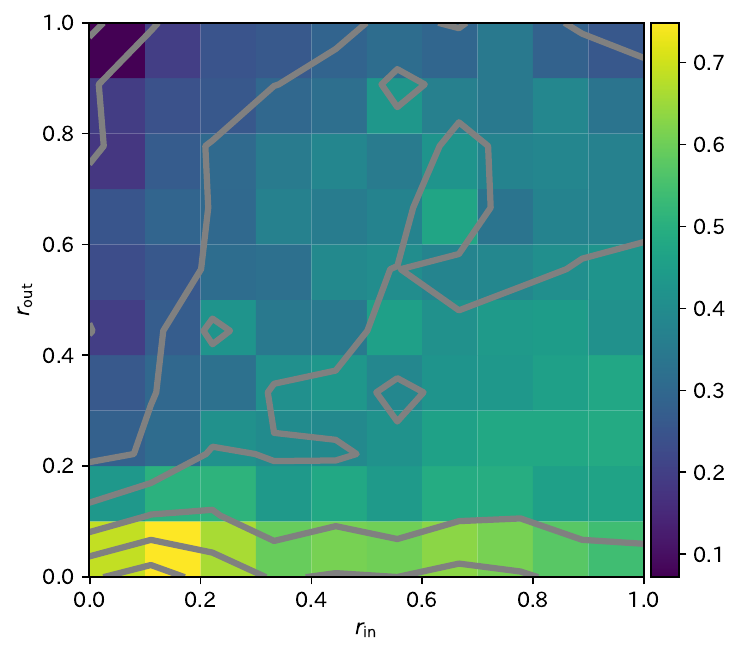}%
        \caption{Percentage of retweets} \label{fig:mutualratio-contour_p_retweets}%
    \end{subfigure}%
    \begin{subfigure}[t]{.24\linewidth}
        \centering%
        \includegraphics[width=.98\linewidth]{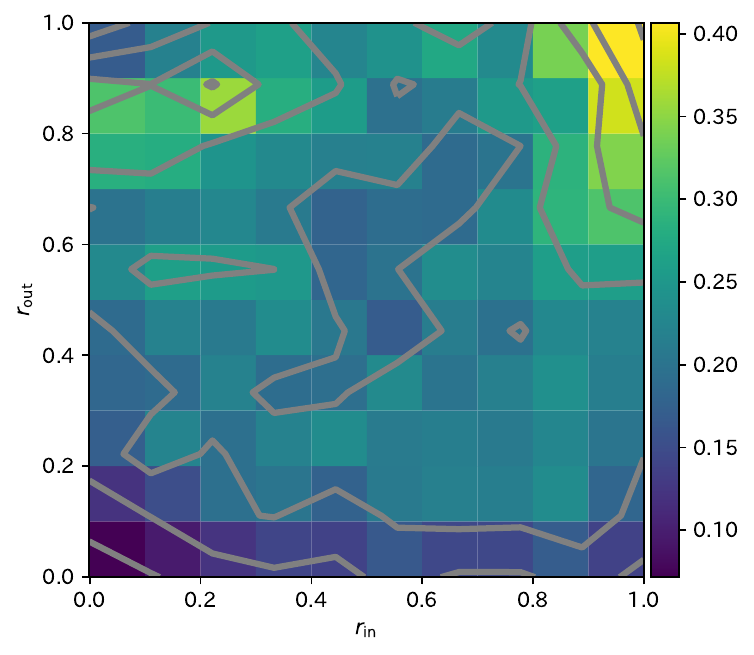}%
        \caption{Percentage of replies} \label{fig:mutualratio-contour_p_replies}%
    \end{subfigure}%
    \begin{subfigure}[t]{.24\linewidth}
        \centering%
        \includegraphics[width=.98\linewidth]{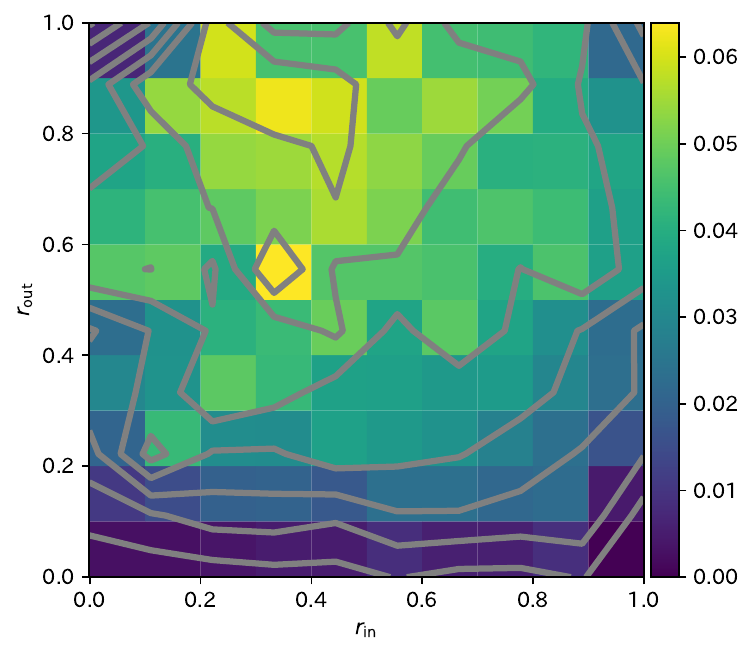}%
        \caption{Percentage of quotes} \label{fig:mutualratio-contour_p_quotes}%
    \end{subfigure}\\
    \begin{subfigure}[t]{.24\linewidth}
        \centering%
        \includegraphics[width=.98\linewidth]{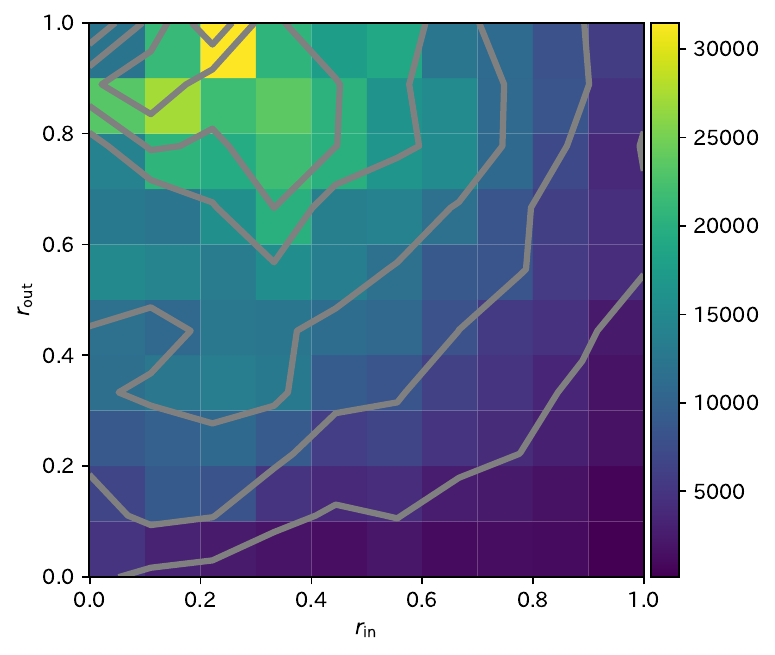}%
        \caption{Number of total posts} \label{fig:mutualratio-contour_statuses_count}%
    \end{subfigure}%
    \begin{subfigure}[t]{.24\linewidth}
        \centering%
        \includegraphics[width=.98\linewidth]{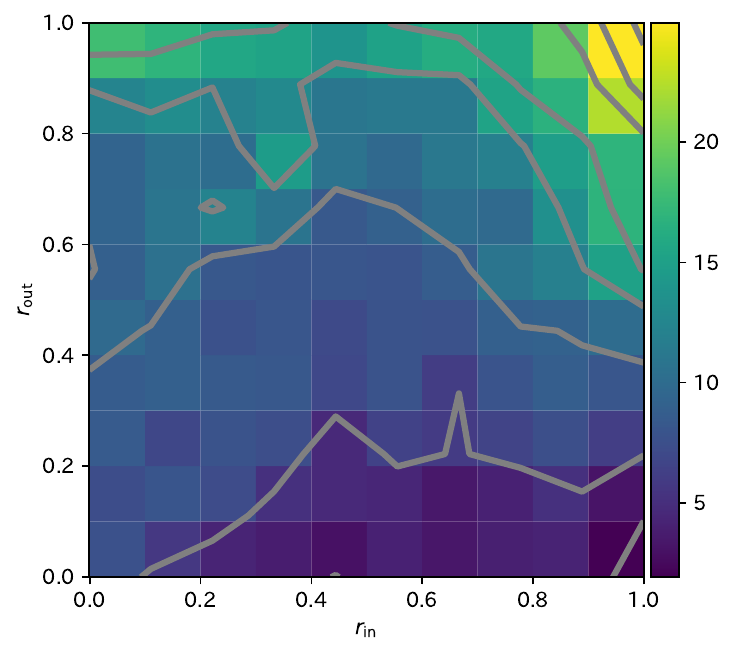}%
        \caption{Tweets per day} \label{fig:mutualratio-contour_tweets_per_day}%
    \end{subfigure}%
    \begin{subfigure}[t]{.24\linewidth}
        \centering%
        \includegraphics[width=.98\linewidth]{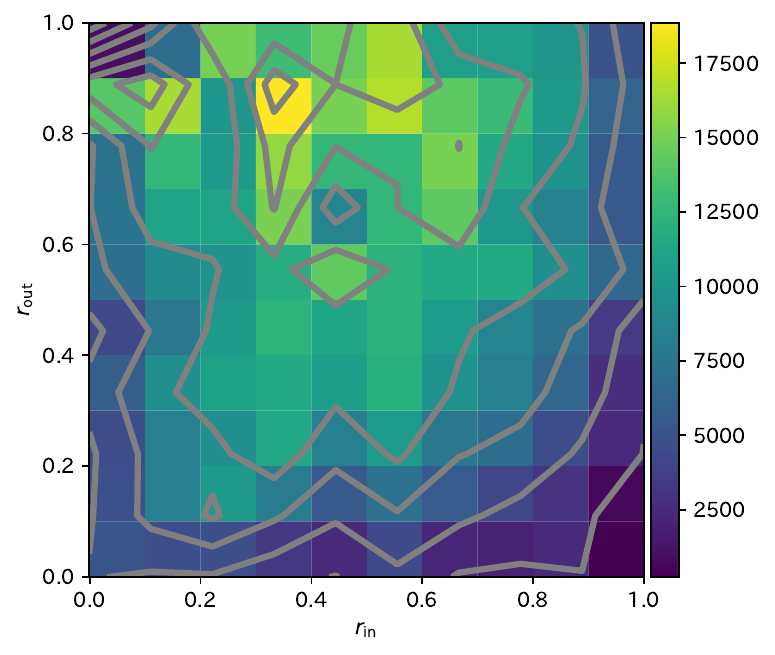}%
        \caption{Number of likes} \label{fig:mutualratio-contour_favourites_count}%
    \end{subfigure}%
    \begin{subfigure}[t]{.24\linewidth}
        \centering%
        \includegraphics[width=.98\linewidth]{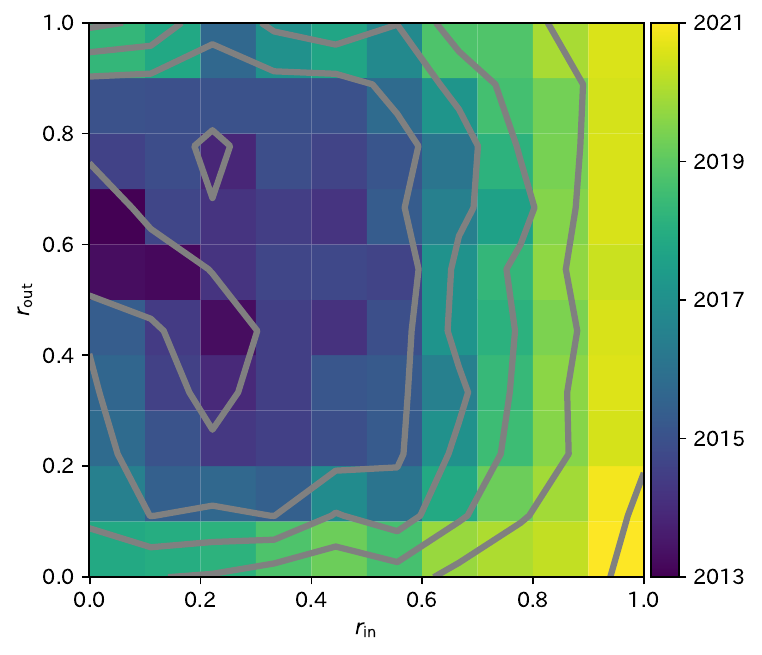}%
        \caption{Account created at} \label{fig:mutualratio-contour_created_at}%
    \end{subfigure}\\
    \begin{subfigure}[t]{.24\linewidth}
        \centering%
        \includegraphics[width=.98\linewidth]{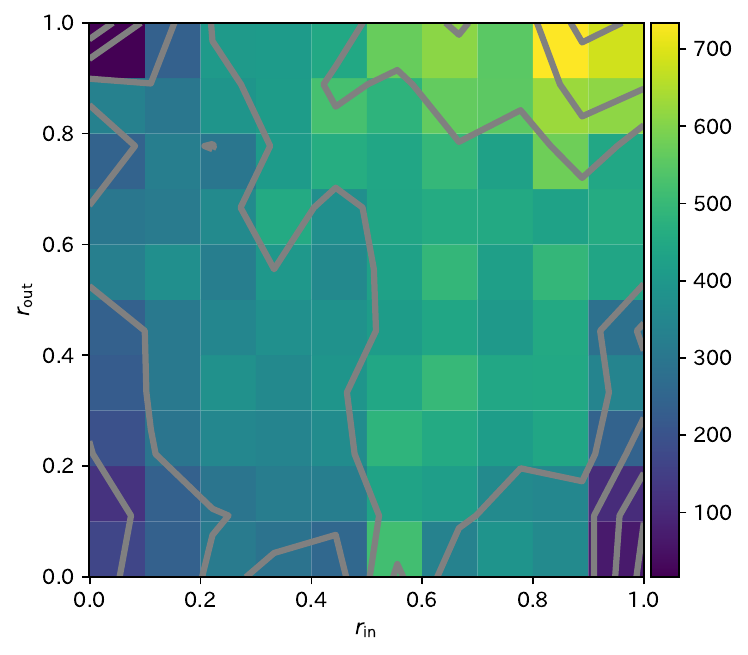}%
        \caption{Number of followees} \label{fig:mutualratio-contour_friends_count}%
    \end{subfigure}%
    \begin{subfigure}[t]{.24\linewidth}
        \centering%
        \includegraphics[width=.98\linewidth]{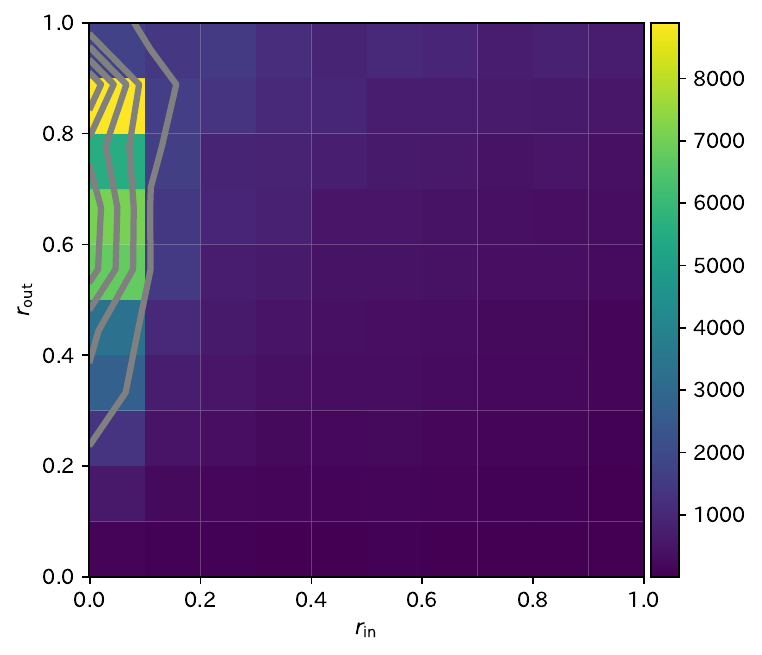}%
        \caption{Number of followers} \label{fig:mutualratio-contour_followers_count}%
    \end{subfigure}\\
    \begin{subfigure}[t]{.24\linewidth}
        \centering
        \includegraphics[width=.98\linewidth]{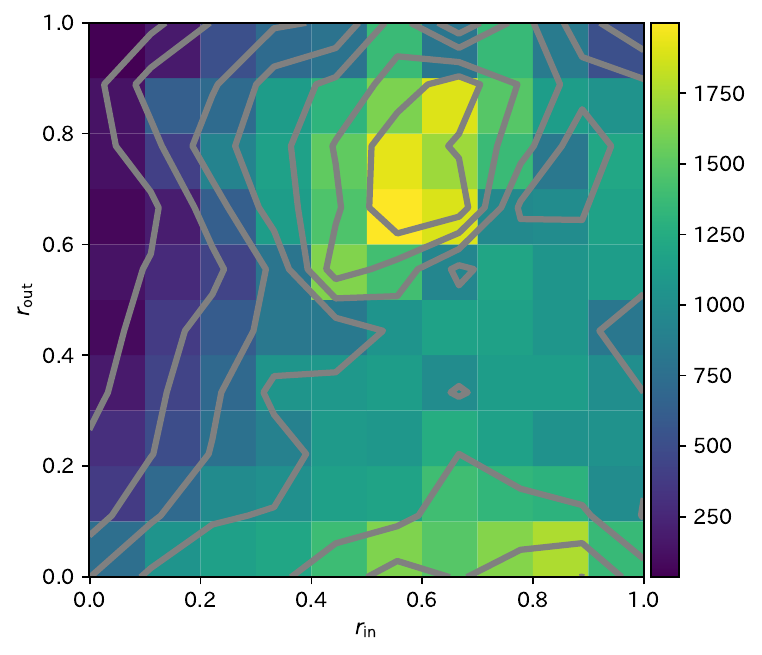}%
        \caption{Mean number of retweeted} \label{fig:mutualratio-contour_mean_retweeted}%
    \end{subfigure}%
    \begin{subfigure}[t]{.24\linewidth}
        \centering
        \includegraphics[width=.98\linewidth]{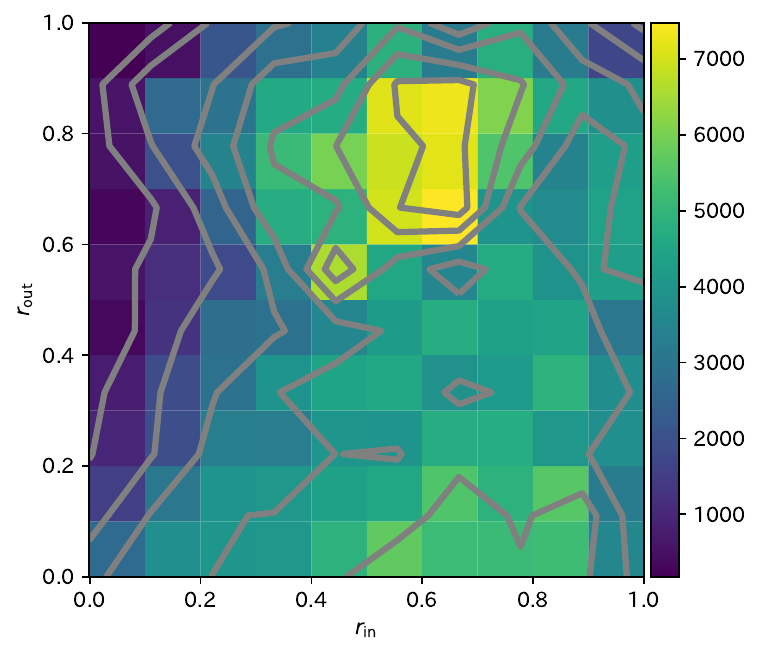}%
        \caption{Mean number of liked} \label{fig:mutualratio-contour_mean_favorited}%
    \end{subfigure}%
    \begin{subfigure}[t]{.24\linewidth}
        \centering
        \includegraphics[width=.98\linewidth]{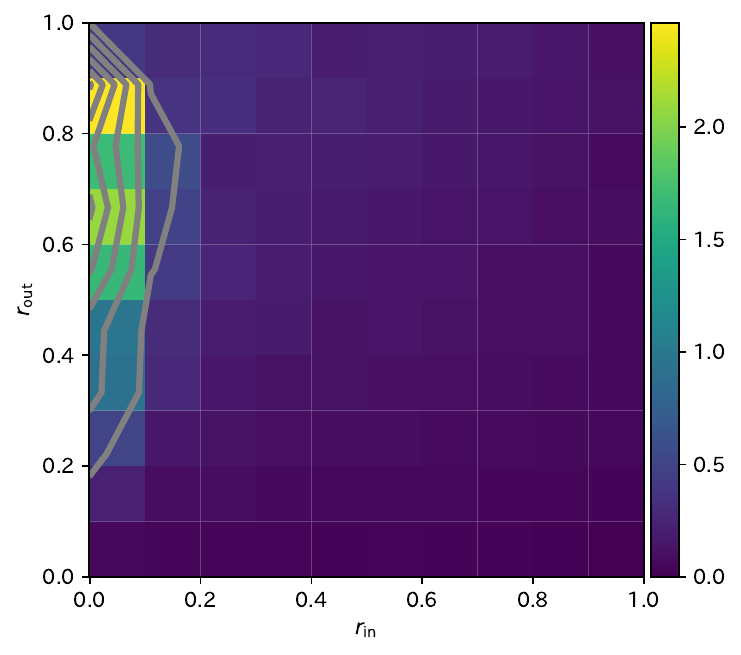}%
        \caption{Mean number of retweeted of original posts} \label{fig:mutualratio-contour_mean_original_retweeted}%
    \end{subfigure}%
    \begin{subfigure}[t]{.24\linewidth}
        \centering
        \includegraphics[width=.98\linewidth]{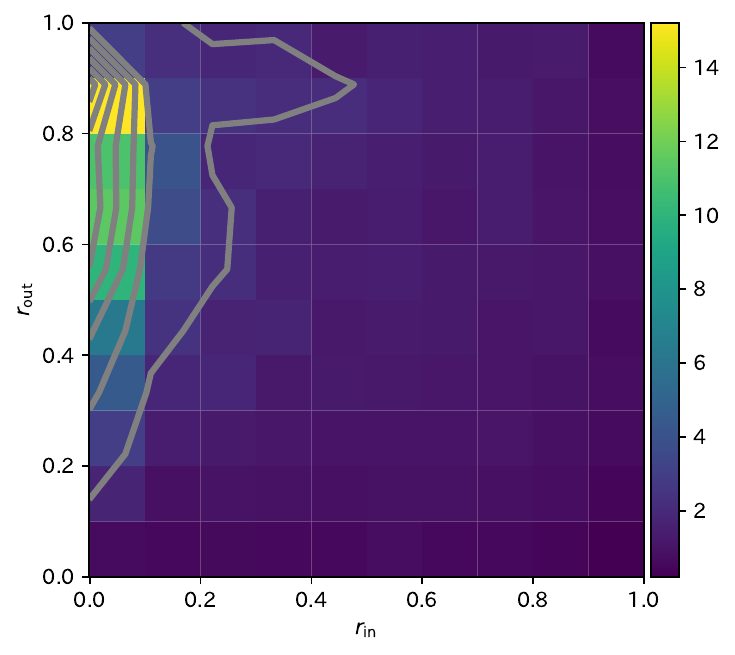}%
        \caption{Mean number of liked of original posts} \label{fig:mutualratio-contour_mean_original_favorited}%
    \end{subfigure}%
    \caption{Continuous behavioral gradients across reciprocity space. Heatmaps showing median user property values across a $10 \times 10$ grid of the two-dimensional reciprocity space ($r_\mathrm{in}$, $r_\mathrm{out}$), with contour lines indicating smooth transitions between behavioral regions. \textit{(Caption continues on the following page.)}}
    \label{fig:user-property-continuous}
\end{figure*}
\begin{figure*}[tp]
    \ContinuedFloat %
    \caption{\textbf{(Continued)} The analysis reveals that user behavior exists along continuous quantitative dimensions rather than discrete categories. (a--d) Tweet composition patterns showing gradual transitions in content creation behavior. (e--h) Activity and temporal patterns demonstrating smooth gradients in user engagement and platform adoption. (i--j) Network structure metrics revealing continuous variation in social connectivity patterns. (k--l) Overall content engagement showing how different reciprocity positions influence content reach. (m--n) Original content engagement metrics, with peak values occurring in an intermediate high engagement zone (approximately $r_\mathrm{in}$ 0.5--0.7, $r_\mathrm{out}$ 0.6--0.9) where balanced reciprocal connections optimize content virality. Each cell represents the median value for users within that reciprocity range.}
\end{figure*}

\section{Limitations and Future Directions}

Our analysis is based on data from a single social media platform (Twitter) collected from English-language posts in the 1\% sample stream during July 2021.
This sampling approach likely overrepresents active users and English speakers while underrepresenting non-English-speaking communities and users who rarely or never post content.
Despite these sampling constraints, we believe that our analysis of 48,830 users provides sufficiently robust results to establish the validity of our reciprocity-based framework.

Our framework relies exclusively on structural network properties to characterize user behavior.
While we demonstrate that these structural features effectively capture patterns in user activity and content engagement, our approach does not incorporate detailed content analysis or contextual information about interaction partners.
Consequently, some nuanced aspects of user behavior and motivation may not be fully reflected in our reciprocity-based characterization.

Future research should investigate how cultural differences influence user positioning within the reciprocity space, analyze user interaction patterns across multiple platforms and time periods, and explore the underlying mechanisms that drive different reciprocity patterns.
Such studies would provide in-depth insights into the generalizability and temporal dynamics of the behavioral patterns we have identified.

\bibliographystyle{elsarticle-num-names}
\bibliography{reference}

\end{document}